\documentclass[12pt,floatfix,groupedaddress,superscriptaddress,longbibliography]{revtex4-1}
\usepackage[english]{babel}
\usepackage{graphicx}
\usepackage{color}

\begin{document}

\title{A new high-temperature quantum spin liquid  with polaron spins}

\author{Martin Klanj\v{s}ek}
\affiliation{Jo\v{z}ef Stefan Institute, Jamova 39, SI-1000 Ljubljana,
Slovenia}

\author{Andrej Zorko}
\affiliation{Jo\v{z}ef Stefan Institute, Jamova 39, SI-1000 Ljubljana,
Slovenia}

\author{Rok \v Zitko}
\affiliation{Jo\v{z}ef Stefan Institute, Jamova 39, SI-1000 Ljubljana,
Slovenia}

\author{Jernej Mravlje}
\affiliation{Jo\v{z}ef Stefan Institute, Jamova 39, SI-1000 Ljubljana,
Slovenia}

\author{Zvonko Jagli\v ci\'{c}}
\affiliation{Faculty of Civil and Geodetic Engineering, University of
Ljubljana, Ljubljana, Slovenia}
\affiliation{Institute of Mathematics, Physics and Mechanics,
Ljubljana, Slovenia}

\author{Pabitra Kumar Biswas}
\affiliation{ISIS Pulsed Neutron and Muon Source, STFC Rutherford
Appleton Laboratory}

\author{Peter Prelov\v sek}
\affiliation{Jo\v{z}ef Stefan Institute, Jamova 39, SI-1000 Ljubljana,
Slovenia}
\affiliation{Faculty of Mathematics and Physics, University of
Ljubljana, Jadranska 19, Ljubljana, Slovenia}

\author{Dragan Mihailovi\v c}
\affiliation{Jo\v{z}ef Stefan Institute, Jamova 39, SI-1000 Ljubljana,
Slovenia}
\affiliation{Faculty of Mathematics and Physics, University of
Ljubljana, Jadranska 19, Ljubljana, Slovenia}

\author{Denis Ar\v con$^{\ast}$}
\affiliation{Jo\v{z}ef Stefan Institute, Jamova 39, SI-1000 Ljubljana,
Slovenia}
\affiliation{Faculty of Mathematics and Physics, University of
Ljubljana, Jadranska 19, Ljubljana, Slovenia}




\begin{abstract}
 The existence of a quantum spin liquid  (QSL) in which  quantum fluctuations of  spins are sufficiently strong to preclude spin ordering down to zero temperature
  was originally proposed theoretically more than 40 years ago, but its experimental realisation turned out to be very elusive. Here we report on an almost ideal spin liquid state that appears to be realized by atomic-cluster spins on the triangular lattice of a charge-density wave (CDW) state of 1T-TaS$_2$. In this system, the charge excitations have a well-defined gap of $\sim 0.3$~eV, while nuclear magnetic quadrupole resonance and muon spin relaxation experiments reveal that the spins show gapless quantum spin liquid dynamics and no  long range magnetic order down to 70~mK. Canonical $T^{2}$ power-law temperature dependence of the  spin relaxation dynamics characteristic of a QSL is observed from 200~K to $T_f= 55$~K. Below this temperature we observe a new gapless state with reduced density of spin excitations and high degree of local disorder signifying new quantum spin order emerging from the QSL. 
\end{abstract}


\maketitle

A resonating valence bond (RVB) state \cite{anderson73} as a new kind of insulator  was proposed to be the ground state of the triangular-lattice $S=1/2$ Heisenberg antiferromagnet instead of a more conventional 
N\'{e}el state to account for the unusual magnetic properties of a perfect triangular atomic lattice of Ta atoms in the layered transition metal dichalcogenide 1T-TaS$_2$. Since then, the list of materials with triangular lattice and with properties indicating 
the existence of a quantum spin-liquid (QSL), i.e., a state without spontaneously broken triangular lattice symmetry and whose behavior is dominated by emergent fractional excitations \cite{Balents}, is still remarkably short: it includes YbMgGaO$_4$ \cite{Shen2016, Paddison2017} and the organic molecular solids $\kappa$-(ET)$_2$Cu$_2$(CN)$_3$ and EtMe$_3$Sb[Pd(dmit)$_2$]$_2$ \cite{Itou_NP, Pratt_Nature2011}.
However, the latter organic molecular solids do not form a perfect triangular lattice implying that they are not ideal model systems. A promising recent example, YbMgGaO$_4$ has a perfect triangular lattice, but the presence of strong spin-orbit coupling makes the nearest-neighbor magnetic interactions anisotropic, again making the system far from an ideal realization of the original idea. Compared to these compounds,  layered dichalcogenides have perfect triangular lattice geometry and a weaker spin-orbit coupling \cite{Rossnagel}, offering a possibility for obtaining a unique insight into the competition between antagonistic QSL and N\'{e}el states \cite{LRO, white}, however, so far no signatures of QSL behaviour have been observed with spins on atomic lattice sites. 

\begin{figure}[htbp]
\includegraphics[width=1.0\linewidth]{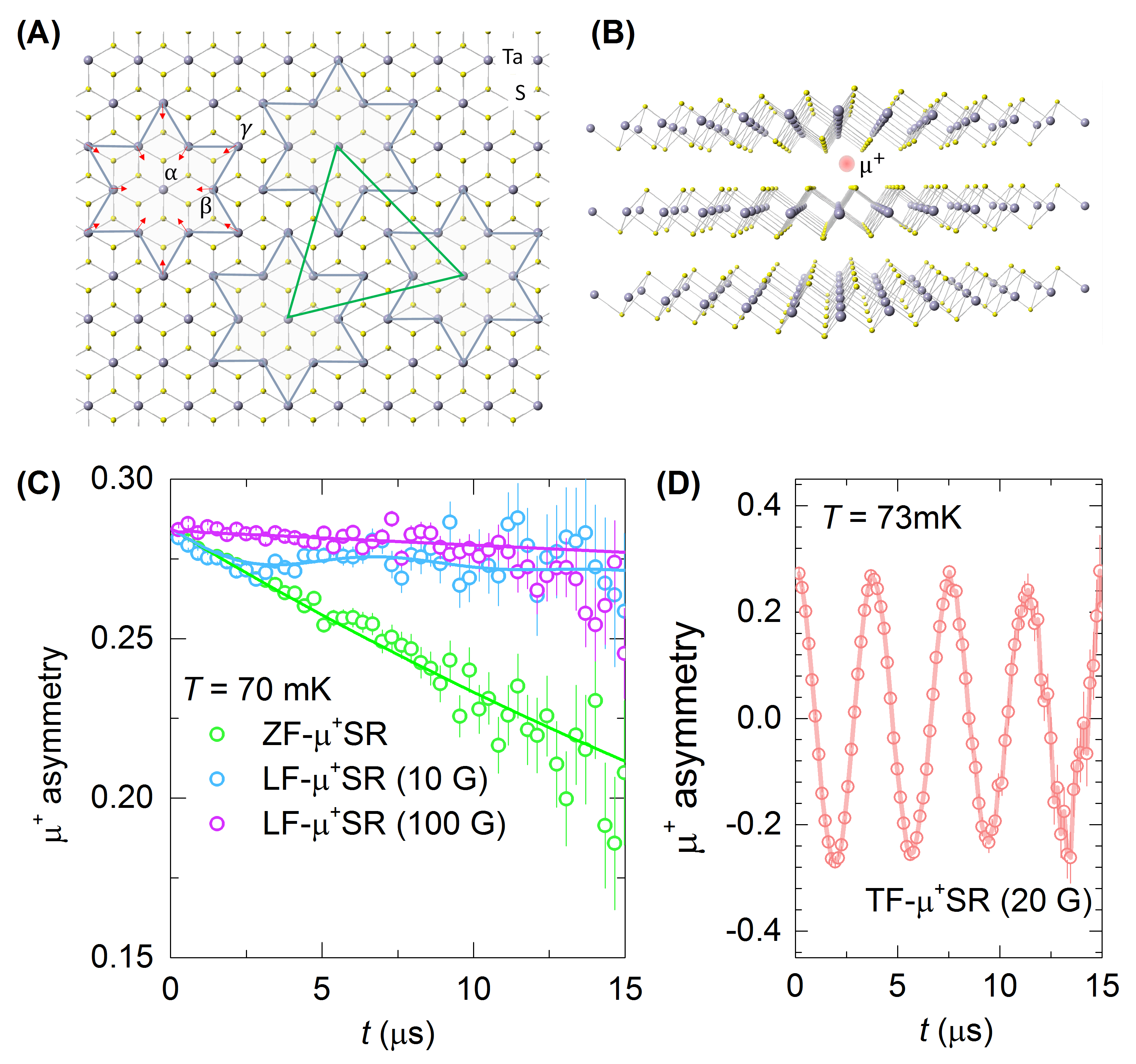}
\caption{\label{fig1} {In the low-temperature commensurate CDW state, the Ta atoms in the TaS$_2$ layer form a characteristic Star-of-David arrangement (solid blue line) leading to three inequivalent Ta sites  --  hereby labeled as $\alpha$, $\beta$ and $\gamma$ sites. Large in-plane Ta atom displacements  are marked by red arrows. The Star-of-David Ta atom clusters form a geometrically frustrated, perfect triangular lattice (solid green line) of polarons. {\bf (B)} The positive muon (red circle) stopping site is most probably close to the negative sulfur ions in the space between TaS$_2$ layers.  {\bf (C)} $\mu^+$SR signal in zero-field (green circles), and in longitudinal-fields of 10 G (blue circles) and 100 G (violet circles)  at $T=70$~mK. {\bf (D)} $\mu^+$ asymmetry in a weak transverse field of 20 G (red circles) at $T=73$~mK. The solid line is a fit to $P(t)=A\cos \left(2\pi\nu_{\rm wTF}t\right)\exp\left(-t/T_{1\mu}\right)$  yielding a $\mu^+$ precession frequency $\nu_{\rm wTF} = 265.4$~kHz and a relaxation time $T_{1\mu}=60(5)\, \mu$s.} 
}
\end{figure}

1T-TaS$_{2}$ is a remarkably versatile model system for  studying a variety of unusual physical phenomena \cite{Rossnagel, Stojchevska177, IwasaSciAdv, DMNatComm}, and is at the same time an experimentally amenable testing ground for new theoretical concepts in correlated electron systems. It's low-temperature state is particularly interesting because it supports a commensurate arrangement of polarons on a triangular lattice.
At high temperatures, the material is a metal, which first undergoes a transition to an incommensurate charge-density wave (IC-CDW) state due to a Fermi surface instability at 545~K. To minimize the strain arising from the IC-CDW and the underlying lattice, this state transforms to a nearly commensurate (NC) and eventually becomes commensurate (C-CDW) below $T_{\rm NC-C}=180$~K on cooling. In this low-temperature phase, the Ta atoms are grouped into 13-atom clusters forming a Star-of-David arrangement with large in-plane Ta atom displacements towards the central Ta atom (Fig. 1A). Experimentally, 1T-TaS$_2$ is insulating in the C-CDW phase \cite{Rossnagel}, which is unexpected, since it has an  odd number of valence electrons per unit cell. This remarkable behaviour does not allow an interpretation of the insulating state in terms of the conventional band theory.
To resolve this puzzle, the Mott-Hubbard mechanism has been invoked \cite{Fazekas_Phil_1979}, 
according to which one electron is localized on each Ta-atom cluster due to electron correlation effects giving rise to a Mott insulating state with $S=1/2$ spins, approximately 1 nm apart, arranged on an ideal triangular-based lattice (Fig. 1A). In this state, charge excitations exhibit a large charge gap $\Delta_{\rm c}\sim0.3$ eV \cite{ARPES, Perfetti2005, optics, STM1, STM2}, while transport anisotropy measurements show that the system remains essentially two-dimensional (2D) \cite{rho_Dragan}. Given that we have an apparently ideal realization of $S=1/2$ spins arranged on a perfect triangular lattice, the question that arises is: what is the spin state of 1T-TaS$_2$? If the material is  a band insulator, a large  gap should be observed also in the spin excitation spectrum. On the other hand, theoretically, within the 2D Heisenberg model of localized spins on nearest-neighbor sites of the triangular lattice, such a spin lattice is known to be magnetically ordered with  coplanar spins oriented at 120 degrees with respect  to one another \cite{LRO, white}. Alternatively,  it may form a new and unusual state of matter, perhaps the elusive QSL, as originally proposed by Anderson \cite{anderson73}, if the spin interactions go beyond the Heisenberg model \cite{triang_PRL, triang_PRB}.

Here we present  the first muon-spin-relaxation ($\mu^+$SR) measurements on 1T-TaS$_2$, which  reveal the spin state over a very broad range of temperatures from 70~mK to 210~K, and measurements of electronic spin fluctuations using nuclear quadrupole resonance (NQR), which reveal a
  gapless  QSL-like behavior in part of the C-CDW phase. Remarkably,  below $\sim 55$~K  a new and unusual spin  state emerges, which does not show any evidence of any kind of magnetic ordering  down to 70~mK. These new findings  not only confirm the  prediction of  QSL  in 1T-TaS$_2$, which has been unresolved for over 40 years, but also raise questions about the nature of emergent states out of QSLs. 


\begin{figure}[htbp]
\includegraphics[width=1.0\linewidth]{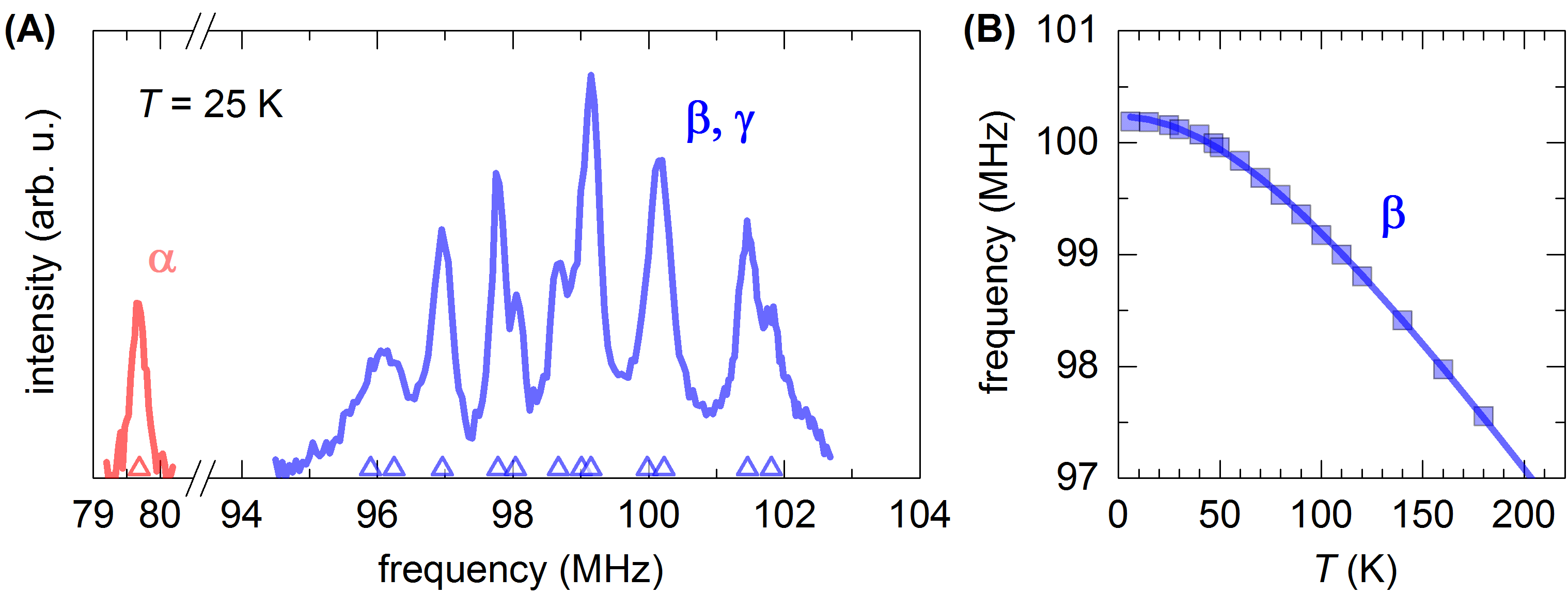}
\caption{\label{fig2} {{\bf (A)} The $^{181}$Ta NQR spectrum measured at $T=25$~K. The isolated peak at the resonance frequency of $\nu\approx 79$~MHz is assigned to the Ta $\alpha$ site whereas the group of 12 overlapping peaks between $\nu=96$ and 102~MHz belongs to the Ta $\beta$ and $\gamma$ sites. Triangles mark the position of individual peaks. {\bf (B)} The temperature dependence of the $\beta$ site $^{181}$Ta NQR frequency (circles) is similar as in layered metals where it is governed by the two-dimensional phonon spectrum. The solid line is a guide to the eye.} 
}
\end{figure}

To probe the potential magnetic order or spin freezing we initially performed 
muon spin relaxation spectroscopy.  The positively charged $\mu^+$ particle is assumed to localize in between the TaS$_2$ layers close to negatively charged S$^{2-}$ ions (Fig. 1B),  from where it extremely sensitively  probes small-moment magnetism, irrespective of whether it is long- or short-ranged,  spatially inhomogeneous or even incommensurate \cite{muSR}. Significantly, in the zero-field (ZF) $\mu^+$SR experiment  conducted at 70~mK (Fig. 1C), we did not observe any  oscillation of the $\mu^+$SR signal that would imply a presence of frozen local  magnetic fields. Further direct evidence ruling out any long-range magnetic order comes from $\mu^+$SR measurements in a weak-transverse-field (wTF) of 2~mT at the same temperature (Fig. 1D).
The local magnetic field at the muon site is the sum of the applied and the internal magnetic fields. In magnetically ordered phases, the latter is typically of the order of 
several tens of ~mT and thus exceeds the small applied wTF. However, the oscillation frequency of the wTF $\mu^+$SR asymmetry, $\nu_{\rm wTF}=265.4$~kHz, corresponds to the expected $\mu^+$ precession frequency in the applied wTF  ($\nu_{\rm wTF}=\gamma_\mu B_{\rm wTF}/2\pi$, $\gamma_\mu/2\pi =135.5$~MHz/T) thus conclusively ruling out any static local field at the $\mu^+$ site. Moreover, the amplitude of the wTF $\mu^+$SR signal corresponds to the full muon asymmetry, thus proving that the absence of any long-range magnetic order in 1T-TaS$_2$ pertains to the bulk of the sample. Complementary longitudinal field (LF) $\mu^+$SR measurements (Fig. 1C, Fig. S1)  show a weak relaxation process due to the dynamics of the local fields of electronic origin. This yields a very small muon relaxation rate $\lambda$ (i.e., $\lambda=0.0023$~$\mu{\rm s}^{-1}$ at 70~mK) with no (or very little) temperature dependence (Fig. S2). 

Given the absence of magnetic order and the fact that the observed spin dynamics is not well resolved in  $\mu^+$SR measurements,  we have further studied the electron spin excitations by another local magnetic probe, the $^{181}$Ta nuclear magnetic moments.
The $^{181}$Ta (nuclear spin $I=7/2$) NQR spectrum of a collection of 1T-TaS$_2$ single crystals measured at $T=25$~K comprises  an isolated  peak centered at 78~MHz and a group of 12 overlapping  peaks in the frequency region spanning from 96 to 102~MHz (Fig. 2A). All peaks are due to the $^{181}$Ta  transitions between the nuclear spin states $m_I=1/2\leftrightarrow 3/2$.  The NQR resonance frequency $\nu$ is proportional to the largest eigenvalue of the electric field gradient (EFG) tensor and is thus an extremely sensitive function of the charge distribution around a given nucleus. Therefore, the individual peaks in the measured NQR spectrum reflect variations of EFG values over the three different Ta sites. The isolated peak at 78~MHz is assigned  to the central Ta $\alpha$ site \cite{TaNQR1, TaNQR2}, whereas the peaks  around 100~MHz are assigned to the peripheral  Ta $\beta$ and $\gamma$ sites (Fig. 1A). 
The fact that this part of the spectrum is fully split into two sets of $6+6$ peaks is consistent with the symmetry and structure of the Ta deformations surrounding the central Ta atom in the star of David cluster and rules out bilayer lattice dimerization where the singlet formation between layers breaks the trigonal symmetry and which would give rise to $3+3$ multiplet of peaks \cite{TaNQR1}. 
A characteristic temperature dependence of the $\alpha$ site $^{181}$Ta resonance frequency, where the dependence gradually changes from the linear at high  to the quadratic at lower temperatures  (Fig. 2B), is understood to arise from the two-dimensional character of the phonon spectrum in layered materials \cite{BorsaPRL}. Importantly, a smooth and monotonic variation of $\nu_\beta(T)$ [and likewise also of $\nu_\alpha(T)$ and $\nu_\gamma(T)$, Fig. S4] proves that the same charge superlattice structure is present for all temperatures below $T_{\rm NC-C}$.

\begin{figure}[htbp]
\includegraphics[width=0.70\linewidth]{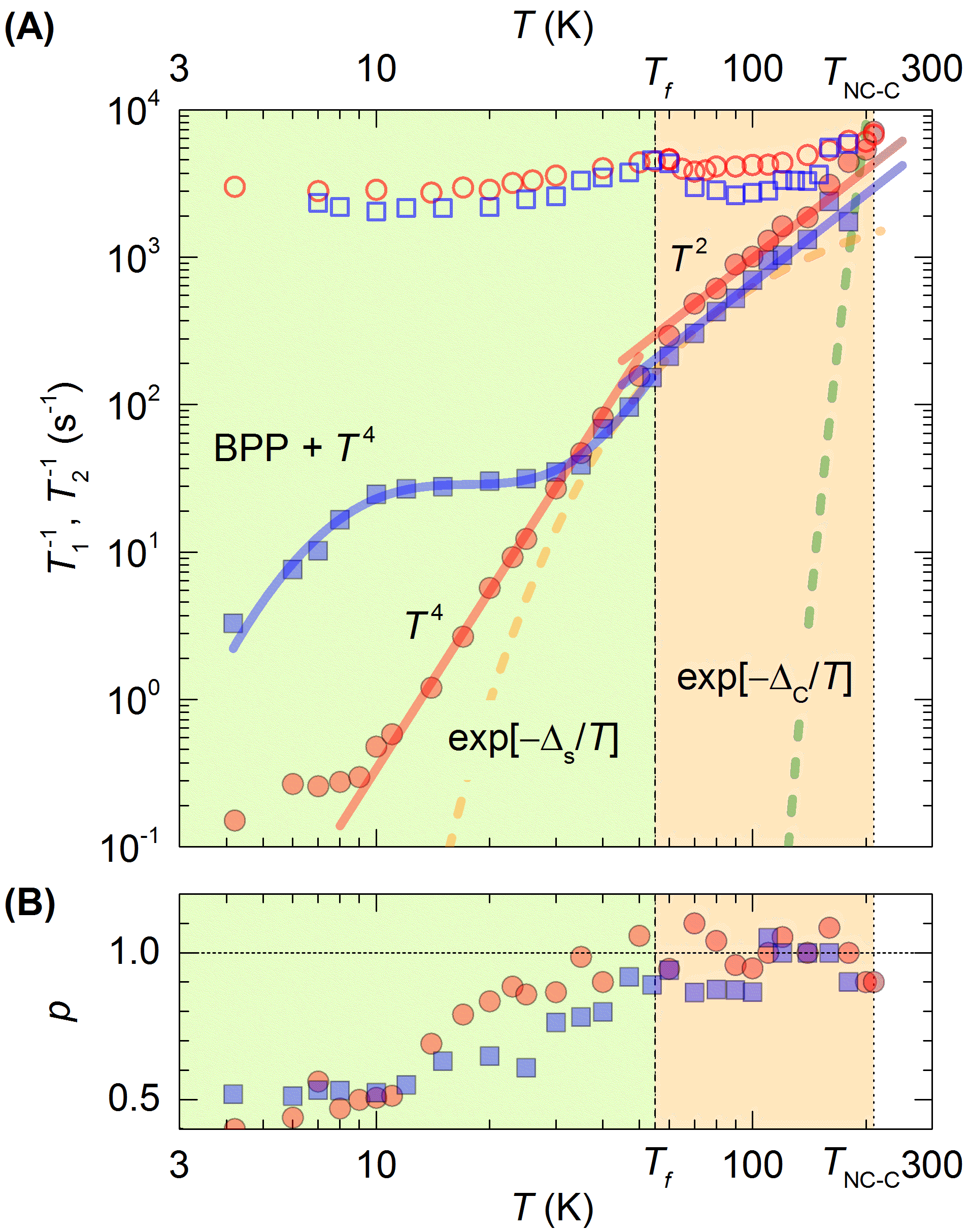}
\caption{\label{fig3} {{\bf (A)} The temperature dependence of the $^{181}$Ta spin-lattice relaxation rates $1/T_1$ for the Ta $\alpha$ site (solid red circles) and for the Ta $\beta$ site (solid blue squares). The $^{181}$Ta spin-spin relaxation rates $1/T_2$ are  represented by open red circles and open blue squares for the Ta $\alpha$ and $\beta$ sites, respectively. The solid red lines show the power-law temperature dependence of $1/T_1$ in different temperature regimes, changing from the high-temperature $1/T_1\propto T^2$ to a steeper $1/T_1\propto T^4$ dependence below $T_f= 55$~K. The Ta $\beta$ site has an additional relaxation channel described by a Bloembergen-Purcell-Pound (BPP)-like relaxation term (BPP) due to the thermally activated Ta-cluster dynamics yielding an activation energy of $E_{\rm a}/k_{\rm B}=18.4$~K and the attempt frequency  $1/\tau_0=2.1$~GHz. Dashed lines represent the thermally activated temperature dependences of $1/T_1$, if governed by charge excitations across the charge gap $\Delta_{\rm c}$ (green) or by singlet-triplet excitations across  the spin gap $\Delta_{\rm s}$ (orange).   {\bf (B)} The temperature dependence of the stretching exponent $p$ for the  Ta $\alpha$ (red circles) and $\beta$ sites (blue squares). A sudden reduction in $p$ below $T_f$ is indicated by a vertical dashed line. Vertical dotted line at  $T_{\rm NC-C}=210$~K marks the transition between C- and NC-CDW states (as measured on warming).} 
}
\end{figure}

In Fig. 3A we show the temperature dependence of the $^{181}$Ta spin-lattice relaxation rate, $1/T_1$, for the Ta $\alpha$ and $\beta$ sites ($\gamma$ sites behave very similarly to the $\beta$ sites). In a wide temperature range, $55-200$~K, the relaxation rate follows what appears to be a power law temperature dependence.
It is of primary importance that the observed dependence  cannot be attributed to   excitations across the charge gap $\Delta_{\rm c}$, which is expected to follow a much steeper $1/T_1\propto \exp \left(-\Delta_{\rm c}/T\right)$ dependence (green line in Fig. 3A). 
Alternatively, one might try to fit the measured dependence by assuming the presence of an independent spin gap $\Delta_{\rm s}$ leading to a thermally activated $1/T_1 \propto \exp(-\Delta_{\rm s}/T)$ temperature dependence. In this scenario the obtained $\Delta_{\rm s} \sim 160$~K (orange line in Fig. 3A) could be  interpreted as either a spin gap emerging from in-plane spin singlet-triplet excitations or from out-of-plane spin excitations. The size of $\Delta_{\rm s}$ is not unreasonable for the 1T-TaS$_2$ out-of-plane charge bandwidth and the effective Coulomb repulsion \cite{Millis}. However, this scenario fits the observed temperature dependence of $1/T_1$ only over a very narrow temperature range between 35 and 70~K. 
Therefore, the presence of $\Delta_{\rm s}$ cannot explain the observed $1/T_1$ data, nor can it explain the ubiquitously observed small and temperature-independent spin susceptibility (Fig. S3).
Thus, dismissing activated behavior arising from the presence of any kind of gap, $1/T_1$ for all Ta sites obeys a convincing power-law dependence $1/T_1\propto T^n$ with two different power-law exponents $n$ in two different temperature regimes. 
At high temperatures, $T>55$~K, all three Ta sites show a similar spin-lattice relaxation with $n=2\pm 0.2$. Such a temperature dependence is, for instance,  found when the nuclear spin-lattice relaxation is dominated by thermal lattice vibrations. However, the estimated phonon-driven relaxation rate $1/T_1^{\rm ph}\approx 0.45$~s$^{-1}$ at $T=200$~K (see electronic supplementary file for details) is by at least 3-4 orders of magnitude too small to account for the measured relaxation rates in 1T-TaS$_2$. Altogether, these arguments rule out relaxation by charge excitations across $\Delta_{\rm c}$, singlet-triplet excitations separated by $\Delta_{\rm s}$ and thermal lattice vibrations, leaving us with the only alternative that the power-law temperature dependences of $1/T_1$ are governed by fluctuations of the electron spins.

Below $T_f=55$~K, the temperature dependence of $1/T_1$ changes, but this change is markedly different on different Ta sites. 
For the $\alpha$ site, a transition to another power-law dependence with $n=4$ is observed. On the other hand, for the $\beta$ and $\gamma$ sites $1/T_1$ first approximately levels off and then starts to decrease rapidly again below $\sim 10$~K. The difference of $1/T_1$ between different Ta sites in this low-temperature regime is really striking, with $1/T_1$ of the $\alpha$ site being by up to two orders of magnitude smaller compared to the other two sites. 
Although any long range magnetic order has been clearly ruled out by $\mu^+$SR, the presence of spin fluctuations  at ${\bf q}=\pm (2\pi/3, 2\pi/3)$, which are reminiscent of the $120^\circ$ spin ordering \cite{Kaneko}, cannot be {\em a priori} neglected. 
However, the expected  broad diffusive peaks in $\chi'' ({\bf q},\omega)$  resulting from such spin correlations  cannot account for the unusually efficient filtering of $q$-dependent spin fluctuations required to explain the variation in $1/T_1$ values between different Ta sites by two orders of magnitude and thus alternative explanations are needed. An alternative possibility is the presence of an additional relaxation channel, for which the characteristic frequency of local field fluctuations crosses the NQR frequency on cooling below $T_f$ and whose symmetry is such, that they mainly affect $^{181}$Ta moments at the periphery of the Ta-cluster (i.e., on $\beta$ and $\gamma$ sites). Therefore, 
the temperature dependence of $1/T_1$ below $T_f$ for the Ta $\beta$ and $\gamma$ sites is empirically modeled (Fig. 3A) by two parallel nuclear spin-lattice relaxation channels: $1/T_1=aT^n+b{\tau_{\rm c}\over 1+(\omega \tau_{\rm c})^2}$. The first term describes the power-law dependence with $n= 4$, i.e., the relaxation channel effectively detected by all three Ta sites. The second term is introduced at the periphery of the Star-of-David Ta-cluster and signifies that the autocorrelation function of local field fluctuations is decaying exponentially with  the correlation time $\tau_{\rm c}$. Assuming that $\tau_{\rm c}$ is thermally activated, i.e., $ \tau_{\rm c}=\tau_0\exp \left[E_{\rm a}/k_{\rm B}T\right]$, we find  agreement with the data yielding an activation energy of $E_{\rm a}/k_{\rm B}=18.4$~K and the attempt frequency of $1/\tau_0=2.1$~GHz. Plausible candidates for such low-frequency fluctuations may be low-frequency librational-like modes of the entire Star-of-David, which need to retain the cluster's symmetry in order to be consistent with the data.

The transition at $T_f\simeq 55$~K is accompanied by a  change in the form of the $^{181}$Ta nuclear magnetization recovery curves. Whereas we find nearly perfect single-exponential recovery curves for $T>T_f$, implying a homogeneous spin system, below $T_f$ we need to employ a stretched-exponential fit $M_z(\tau)\propto \exp \left[-\left(\tau /T_1\right)^p\right]$ (here, $M_z(\tau)$ is the $z$-component of the $^{181}$Ta nuclear magnetization, Fig. S5) with a stretching exponent $p$ significantly smaller than $1$ (Fig. 3B). Measurements with $p<1$ suggest the growth of the local-field inhomogeneities at Ta sites leading to a very broad distribution of $1/T_1$ values below $T_f$. 
The observed change of the $1/T_1$ dependence also coincides with the anomaly in the $^{181}$Ta spin-spin relaxation rate $1/T_2$ (Fig. 3A, Fig. S6), which shows a clear maximum at $T_f$ for all the Ta sites. Whereas $1/T_2$ matches $1/T_1$ at 200 K, it is very weakly temperature dependent and 2-3 orders of magnitude larger than $1/T_1$ below $T_f$. Since $1/T_2$ probes local field fluctuations with much lower frequencies than $1/T_1$, we conclude that the observed maximum in $1/T_2$ is a signature of critical behaviour in the low-frequency part $(\omega\rightarrow 0)$ of the local-field dynamics associated with the transition to the low-temperature emergent state.


The central finding of this work is that the magnitude and the temperature dependence of the $^{181}$Ta NQR relaxation rates can only be described by the fluctuations of electron spins within a state that withstands long-range magnetic ordering and whose absence down to 70~mK is unambiguous from the $\mu^+$SR data.
The first and most obvious clue about the nature of the spin state in the range $T_f<T<T_{\rm NC-C}$ comes from the experimental findings revealing that the system is magnetically homogeneous ($p=1$ indicates that the relaxation is the same on all Ta sites), and that the relaxation rate shows a convincing power-law temperature dependence with $n=2\pm 0.2$. 
These findings have important implications for the triangular lattice of the Star-of-David Ta-atom clusters for which QSL emerges as the most probable low-energy state. 
The relaxation rate $1/T_1\propto T\int \left(\chi'' (q,\omega)/\omega\right) dq\propto T^n$ with  $n=2$ means that the $q$-integrated imaginary part of dynamic spin susceptibility $\chi'' (q,\omega)$ is linear in temperature. For gapless spin liquids with a spinon Fermi surface we expect $1/T_1T=const$ \cite{Itou_NP}, which is inconsistent with the experimental $n=2$. In the case of a fully opened spin gap, $1/T_1T$ should decay exponentially with temperatures, which is also inconsistent with the experiment. Therefore, the only candidate among the quantum spin liquids  that fits the present experimental data is the one where the  spin excitations have nodes in $q$-space, and whose energy dependence of the spinon density of states is such that it gives $n=2$. Such power-law dependences with  $n =2$ were encountered previously in several  QSL systems, such as the organic triangular-lattice antiferromagnets \cite{Itou_NP, Pratt_Nature2011} and the disordered triangular lattice systems Sc$_2$Ga$_2$CuO$_7$ \cite{Baenitz_PRB2016}.

\begin{figure}[htbp]
\includegraphics[width=1.0\linewidth]{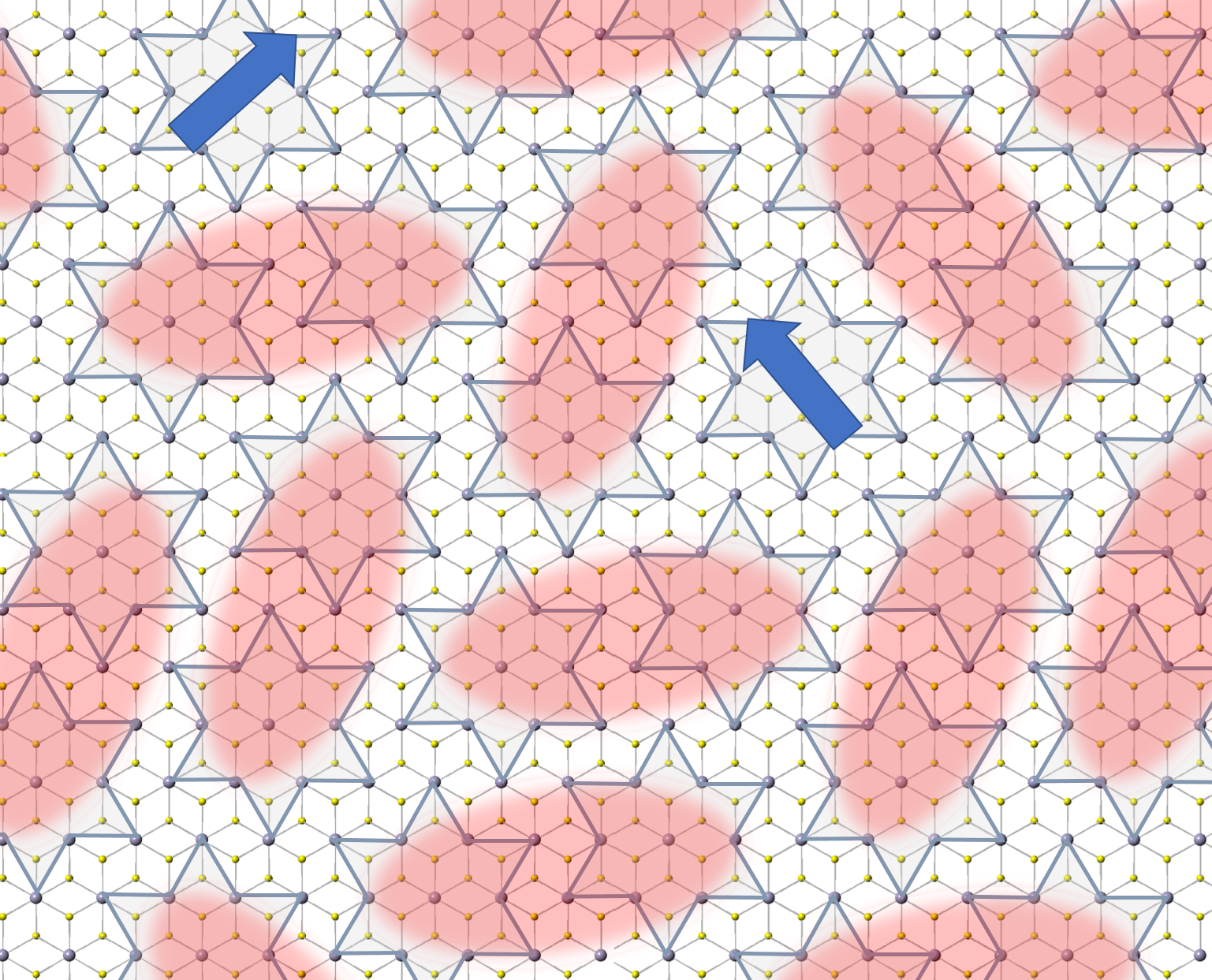}
\caption{\label{fig4} {A schematic illustration of the low-temperature ($T<T_f$)  phase in 1T-TaS$_2$ with the arrangement of the Star-of-David singlets (red shaded areas) in a spatially random manner. This state still exhibits a gapless behavior for the low energy fractional excitations (blue arrows).} 
}
\end{figure}

Below $T_f=55$~K, the QSL state of 1T-TaS$_2$ appears to  support the emergence of a new exotic low-temperature  state characterized by:
\begin{itemize}
\item[(i)]	a broad distribution of $1/T_1$ values (Fig. 3B) where the stretching exponent $p\approx 0.5$ implies a highly inhomogeneous magnetic phase at {\em{all}} Ta sites, 
\item[(ii)]	a slowing down of the spin fluctuations leading to an unusually high power-law exponent $n = 4$ of the $^{181}$Ta relaxation  (Fig. 3A), which implies a suppression of spinon density of states, and
\item[(iii)]	symmetry breaking of the homogeneous spin structure of the Star-of-David Ta-atom clusters, in which the central $\alpha$ site Ta atom shows very different relaxation than the surrounding 12 atoms  (Fig. 3A).
\end{itemize}
The extremely broad distribution of the $^{181}$Ta relaxation rates is a compelling evidence for the growing randomness in the spin system as temperature decreases below $T_f$. 
If the out-of-plane hopping were important, we can imagine that spin chains of stacked Star-of-David Ta clusters may form along the crystallographic $c$-axis. Stacking disorder would then introduce randomness into the quasi-1D spin chain. Indeed, 
similar values of $p$ have been recently encountered in the random Heisenberg chain compound BaCu$_2$(Si$_{1-x}$Ge$_x$)$_2$O$_7$ \cite{Shiroka}, which is known to adopt the random-singlet state \cite{Fisher}. However, in a one-dimensional limit, the nuclear $1/T_1$ relaxation rate, which can be elegantly treated within the Tomonaga-Luttinger-liquid formalism \cite{Klanj1, Klanj2},  shows a power-law temperature dependence with $n\leq 3/2$ (for example, $n=1$ in  BaCu$_2$(Si$_{1-x}$Ge$_x$)$_2$O$_7$ \cite{Shiroka}).  In other words, the large exponent $n=4$ is not consistent with such one-dimensional spin interactions and thus the system also for $T<T_f$  remains 2D as 1D out-of-plane spin order can be excluded.
The observed slowing down of spin fluctuations is consistent with the freezing of singlets, with $p\approx0.5$ suggesting a random 2D arrangement shown schematically in Fig. 4. 
 This phase differs from canonical spin glasses in one important aspect: the average value of the spin is zero for each Ta-cluster, $\langle S_i\rangle$, which explains our $\mu^+$SR data. The frozen in-plane singlets lead to a spinon ``pseudogap'' with a large slope of the spinon density of states near the spinon Fermi energy, consistent with $n=4$. 
 Finally, the commonly observed very small paramagnetic  spin susceptibility (Fig. S3) with a weak Curie-like component is also consistent with such a state and may even explain the generation of a small fraction of free spins at low temperatures, but the large exchange interaction between them derived from the susceptibility of $J\sim0.13$ eV remains anomalous (see Supplemental Material for discussion of this estimate). 
 At this point, the source of randomness is unclear, but either stacking disorder of TaS$_2$ planes, or slight off-stoichiometry of the sample are likely candidates.

The present study reveals the physics behind a more than 40 years old mystery of the magnetic properties of the low-temperature spin state in 1T-TaS$_2$ and broadly the physics behind the quantum spin liquids on a triangular lattice. The geometrical frustration on the  antiferromagnetic triangular lattice of Star-of-David $S=1/2$ spins is responsible for the formation of a  QSL state below $T_{\rm NC-C}$, which is stable over a remarkably large temperature range and implies a surprisingly large exchange interaction. Out of the QSL, a novel quantum phase with amorphous tiling of frozen singlets emerges at $T_f=55$~K, extending down to 70 mK. The properties of this state are highly unusual. While gapless, it shows indications of a pseudogap in the spinon density of states, with a high-degree of local disorder, yet for each Ta-atom cluster the average value of the spin is zero.  
The origin of the emergent state is proposed to be due to freezing of singlets in the QSL, where the value of $T_{f}$ is determined by the distribution of exchange interactions between Star-of-David spins. It would be interesting to see if such a low-temperature state is generic in other triangular lattices where disorder, next-nearest interactions and/or magneto-elastic effects are present.


\begin{acknowledgments}
Authors acknowledge fruitful discussions with Pietro
Carretta. Authors also acknowledge the contribution of
Petra \v Sutar in sample preparation and structural characterization. D.A. acknowledges the financial support by
the Slovenian Research Agency, grant No. N1-0052, D.M. acknowledges funding by the ERC AdG Trajectory.
\end{acknowledgments}


%


\end{document}